\documentclass[aps,prl,reprint,twocolumn,showpacs,superscriptaddress,floatfix]{revtex4-2}

\usepackage{amsmath,amssymb}
\usepackage{graphicx}
\usepackage{placeins}

\usepackage[colorlinks=true,citecolor=blue,linkcolor=blue,urlcolor=blue]{hyperref}

\makeatletter
\def\fnum@figure{\textbf{\figurename~\thefigure}}
\let\frontmatter@footnote@produce\frontmatter@footnote@produce@footnote
\makeatother

\begin{document}

\title{Anomalous thermal transport in isotope-enriched germanium carbide}

\author{Wenjiang Zhou}
\affiliation{School of Mechanics and Engineering Science, Peking University,
  Beijing 100871, China}

\author{Bai Song}
\email{songbai@pku.edu.cn}
\affiliation{School of Mechanics and Engineering Science, Peking University,
  Beijing 100871, China}
\affiliation{National Key Laboratory of Advanced Micro and Nano Manufacture
  Technology, Peking University, Beijing 100871, China}

\date{\today}

\begin{abstract}
Advances in modern electronics demand the dissipation of heat fluxes approaching $1000~\mathrm{W\,cm^{-2}}$, which drives an urgent need for materials with thermal conductivity ($\kappa$) exceeding $1000~\mathrm{W\,m^{-1}\,K^{-1}}$. 
Here, by solving the Boltzmann transport equation considering both three- and four-phonon interactions, we observe that isotopically pure cubic germanium carbide (GeC) exhibits a remarkable $\kappa$ of $1018~\mathrm{W\,m^{-1}\,K^{-1}}$ at room temperature, which is further corroborated by state-of-the-art machine-learning molecular dynamics simulations. Two unique mechanisms underlie this anomalously high $\kappa$: a remarkably strong isotopic enhancement effect ($289\%$) and a surprisingly modest suppression due to four-phonon scattering ($22\%$)---despite the presence of a wide acoustic–optical phonon gap. 
Moreover, we find that GeC defies the conventional Leibfried-Schl\"omann formula and exhibits a non-monotonic variation of $\kappa$ with external hydrostatic pressure at elevated temperatures. Our findings establish isotope-enriched GeC as a material of ultrahigh $\kappa$ and an ideal platform for exploring exotic phonon physics.
\end{abstract}

\maketitle

The surge in power density has pushed heat fluxes in modern electronics toward $1000~\mathrm{W\,cm^{-2}}$, which severely compromise device performance and reliability in diverse fields such as artificial intelligence, wireless communication, and sustainable energy~\cite{Ball2012,Garimella2012,Waldrop2016,Wu2025,Zhou2026Twist}. 
To overcome this thermal bottleneck and ensure effective heat dissipation, materials with ultrahigh thermal conductivity ($\kappa > 1000~\mathrm{W\,m^{-1}\,K^{-1}}$) are urgently required. However, only a handful of materials reach this regime. For half a century, the search for high $\kappa$ has been primarily guided by Slack's classical criteria~\cite{Slack1973} of light constituent atoms, strong bonding, simple lattice, and weak anharmonicity. The most prominent materials satisfying these requirements are diamond and isotopically enriched cubic boron nitride (cBN), with room-temperature $\kappa$ values of over $2000~\mathrm{W\,m^{-1}\,K^{-1}}$~\cite{Onn1992,Wei1993} and $1600~\mathrm{W\,m^{-1}\,K^{-1}}$~\cite{Chen2020}, respectively. In addition, single-crystal graphite can also exhibit a high $\kappa$ up to around $2000~\mathrm{W\,m^{-1}\,K^{-1}}$, albeit strictly within the basal plane~\cite{Yang2025}.

An alternative route to ultrahigh-$\kappa$ materials is phonon band engineering~\cite{Qian2021}, in which phonon dispersion is tailored to restrict the scattering phase space dictated by energy and momentum conservation. A prime example is cubic boron arsenide (BAs), where the large mass contrast between B and As atoms opens a wide acoustic-optical (a-o) phonon gap, thereby forbidding three-phonon (3ph) interactions between acoustic and optical modes. Meanwhile, the bunching of acoustic branches constrains scattering among three acoustic phonons. With these major scattering channels suppressed, first-principles calculations considering only 3ph interactions reported an unusually high $\kappa$ of $2240~\mathrm{W\,m^{-1}\,K^{-1}}$ for BAs~\cite{Lindsay2013}. Nevertheless, by involving one extra phonon, acoustic-optical coupling can be restored and subsequent calculations revealed that such four-phonon (4ph) processes reduce $\kappa$ by $36\%$-$56\%$ at room temperature~\cite{Feng2017,Yang2019,Chen2020}. Experimental measurements of BAs single crystals yielded a $\kappa$ of about $1200~\mathrm{W\,m^{-1}\,K^{-1}}$~\cite{Tian2018,Li2018,Kang2018}, which is indeed ultrahigh but notably below the 3ph prediction. The case of $\theta$-phase tantalum nitride ($\theta$-TaN) is similar, where a large a-o phonon gap together with strong acoustic bunching and weak phonon-electron scattering yields $\kappa$ values of 995 and $820~\mathrm{W\,m^{-1}\,K^{-1}}$ along the two principal axes. Here, 4ph scatterings again reduce $\kappa$ by about $43\%$~\cite{Kundu2021, li2026metallic}. Additionally, several other materials exhibiting large phonon band gaps and high $\kappa$ values have been theoretically predicted, including cubic germanium carbide (GeC), although previous study focused exclusively on 3ph scatterings~\cite{Lindsay2013_isotope}.

In this work, by combining first-principles lattice dynamics calculations and machine-learning molecular dynamics (MD) simulations, we revisit GeC and report a room-temperature lattice thermal conductivity of $1018~\mathrm{W\,m^{-1}\,K^{-1}}$ upon isotope enrichment, even with higher-order intrinsic phonon scatterings incorporated. 
This anomalously high $\kappa$ arises from two distinct mechanisms: a strong isotopic enhancement of $289\%$ and a modest 4ph suppression of only $22\%$---despite the presence of a wide a-o gap similar to BAs and $\theta$-TaN. 
Our scattering channel-resolved analysis traces this anomaly to selection-rule-enhanced 3ph scatterings which is rooted in the unique structure of the phonon bands. In particular, the flat optical bands of GeC resonantly amplify 3ph interactions between acoustic and optical modes, while the relatively weak bunching of acoustic branches sustains intense three-acoustic-phonon scatterings. In addition, we investigate the effect of hydrostatic pressure on the $\kappa$ of isotopically enriched GeC. At elevated temperatures, we observe an anomalous non-monotonic dependence that defies the conventional Leibfried-Schl\"omann formula.

Within the framework of linearized phonon Boltzmann transport equation (BTE), the lattice thermal conductivity is expressed as
\begin{equation}
\kappa^{\alpha\beta} = \frac{1}{8\pi^{3}} \int_{\mathrm{BZ}} \sum_{p} C_{p\mathbf{q}}\, v_{p\mathbf{q}}^{\alpha}\, F_{p\mathbf{q}}^{\beta}\, d^{3}\mathbf{q},
\label{eq:kappa}
\end{equation}
where $\omega_{p\mathbf{q}}$, $v_{p\mathbf{q}}^{\alpha}$, $C_{p\mathbf{q}}$, and $F_{p\mathbf{q}}^{\beta}$ are the frequency, group velocity, heat capacity, and mean free displacement of the phonon mode with branch index $p$ and wavevector $\mathbf{q}$. The mean free displacement is obtained iteratively as $F_{p\mathbf{q}}^{\beta} = \tau_{p\mathbf{q}}(v_{p\mathbf{q}}^{\beta} + \Delta_{p\mathbf{q}}^{\beta})$, with $\Delta_{p\mathbf{q}}^{\beta}$ quantifying the deviation of the nonequilibrium phonon population from the single-mode relaxation-time approximation. The total scattering rate $1/\tau_{p\mathbf{q}} = 1/\tau_{p\mathbf{q}}^{\mathrm{3ph}} + 1/\tau_{p\mathbf{q}}^{\mathrm{4ph}} + 1/\tau_{p\mathbf{q}}^{\mathrm{ph\text{-}iso}}$ includes 3ph, 4ph, and phonon-isotope contributions~\cite{Tamura1983, broido2007intrinsic, Feng2017}, with their explicit expressions given in Refs.~\cite{Li2014_ShengBTE,Han2022}.

We employ the stochastically temperature dependent effective potential (sTDEP) method~\cite{Hellman2013a,Hellman2013b} as implemented in the TDEP package \cite{Knoop2024} to compute the interatomic force constants at finite temperature, which are a prerequisite for evaluating Eq.~(1). Within the sTDEP iterative scheme, perturbed structures of an $8 \times 8 \times 8$ supercell (1024 atoms) are first generated by sampling atomic displacements from their thermal distribution determined by the second-order force constants at the target temperature. The reference forces for these structures are then computed within density functional theory (DFT) using the Vienna Ab initio Simulation Package (VASP)~\cite{Kresse1996,Kresse1999}, with the projector-augmented wave method~\cite{Blochl1994} and the Perdew-Zunger exchange-correlation functional~\cite{Perdew1981}. A plane-wave cutoff energy of 600~eV and an energy convergence threshold of $10^{-8}$~eV are chosen. A $\Gamma$-centered $1 \times 1 \times 1$ $\mathbf{k}$-point mesh is used. Subsequently, a new set of force constants are extracted by minimizing the difference between the DFT forces and the model-Hamiltonian forces over all perturbed structures. The cutoff radii in the TDEP package \cite{Knoop2024} for the second-order, third-order, and fourth-order force constants are set to 7.7~\AA, 6.0~\AA, and 3.5~\AA, respectively. This procedure is iterated at least three times to ensure convergence. To account for the long-range Coulomb interactions and the longitudinal and transverse optical (LO-TO) phonon splitting~\cite{Gonze1994,Gonze1997}, we employ density-functional perturbation theory (DFPT)~\cite{Baroni2001} to calculate the Born effective charges and the high-frequency dielectric constant. The calculated values at different pressures are listed in Table~S1 of the Supplemental Material~\cite{SM}. Finally, the phonon scattering rates and thermal conductivity are evaluated on a $36 \times 36 \times 36$ $\mathbf{q}$-grid using the FourPhonon package (Table S2)~\cite{Li2014_ShengBTE,Han2022}, with the Gaussian-smearing {\it scalebroad} parameters set to 1.0 and 0.1 for the 3ph and 4ph calculations, respectively~\cite{Zhou2022_SnSe}.

To independently verify our findings, we perform state-of-the-art machine-learning MD, which provides a non-perturbative treatment of phonon scattering to all orders~\cite{Zhou2025superionic,Zhou2026LiNbO3}. A neuroevolution potential (NEP) ~\cite{Fan2021_NEP,Dong2024,Xu2025} is trained on first-principles data through an active-learning scheme, as detailed in the Supplemental Material~\cite{SM}. The thermal conductivity is evaluated with the homogeneous non-equilibrium molecular dynamics (HNEMD) method~\cite{Evans1982,Fan2019HNEMD}, in which an external driving force $\mathbf{F}_i^{\rm ext} = \mathbf{F}_{\rm e} \cdot \mathbf{W}_i$ acts on each atom $i$, with $\mathbf{F}_{\rm e}$ the driving-force parameter and $\mathbf{W}_i = \sum_{j\neq i} \mathbf{r}_{ij} \otimes \partial U_j / \partial \mathbf{r}_{ji}$ the virial tensor. The thermal conductivity tensor follows from
\begin{equation}
\frac{\langle J_\alpha \rangle_{\rm ne}}{TV} = \sum_\beta \kappa^{\alpha\beta} F_{\rm e}^\beta,
\end{equation}
where $\mathbf{J} = \sum_i \mathbf{W}_i \cdot \mathbf{v}_i$ is the heat current, $\mathbf{v}_{i}$ is the velocity of atom $i$, $T$ is the absolute temperature, and $V$ is the volume. The simulations use a 13\,824-atom supercell, and the thermal conductivity is evaluated from MD production runs in the $NVT$ ensemble with a global Nos\'e-Hoover chain (NHC) thermostat~\cite{Martyna1992}, with computational details given in the Supplemental Material~\cite{SM}. Since the machine-learning potential (MLP) carries a finite force error ($\sigma_{\mathrm{mlp}}$) relative to the DFT forces, which scatters phonons in a way analogous to an isotope effect, a correction is required to obtain the intrinsic thermal conductivity. Based on our previous works~\cite{zhou2025insight,Wu2024}, we add controlled force noise ($\sigma_{\mathrm{L}}$) through a local Langevin thermostat~\cite{Bussi2007}, so that the total force error becomes $\sigma_{\mathrm{total}} = \sqrt{\sigma_{\mathrm{mlp}}^2 + \sigma_{\mathrm{L}}^2}$. We then perform a fitting as, $1/\kappa = 1/\kappa_0 + \alpha \sigma_{\mathrm{total}} + \beta \sigma_{\mathrm{total}}^2$ ($\alpha$ and $\beta$ being fitting parameters), and extrapolate to zero force error to obtain the intrinsic thermal conductivity $\kappa_0$.

\begin{figure}[t]
\centering
\includegraphics[width=\columnwidth]{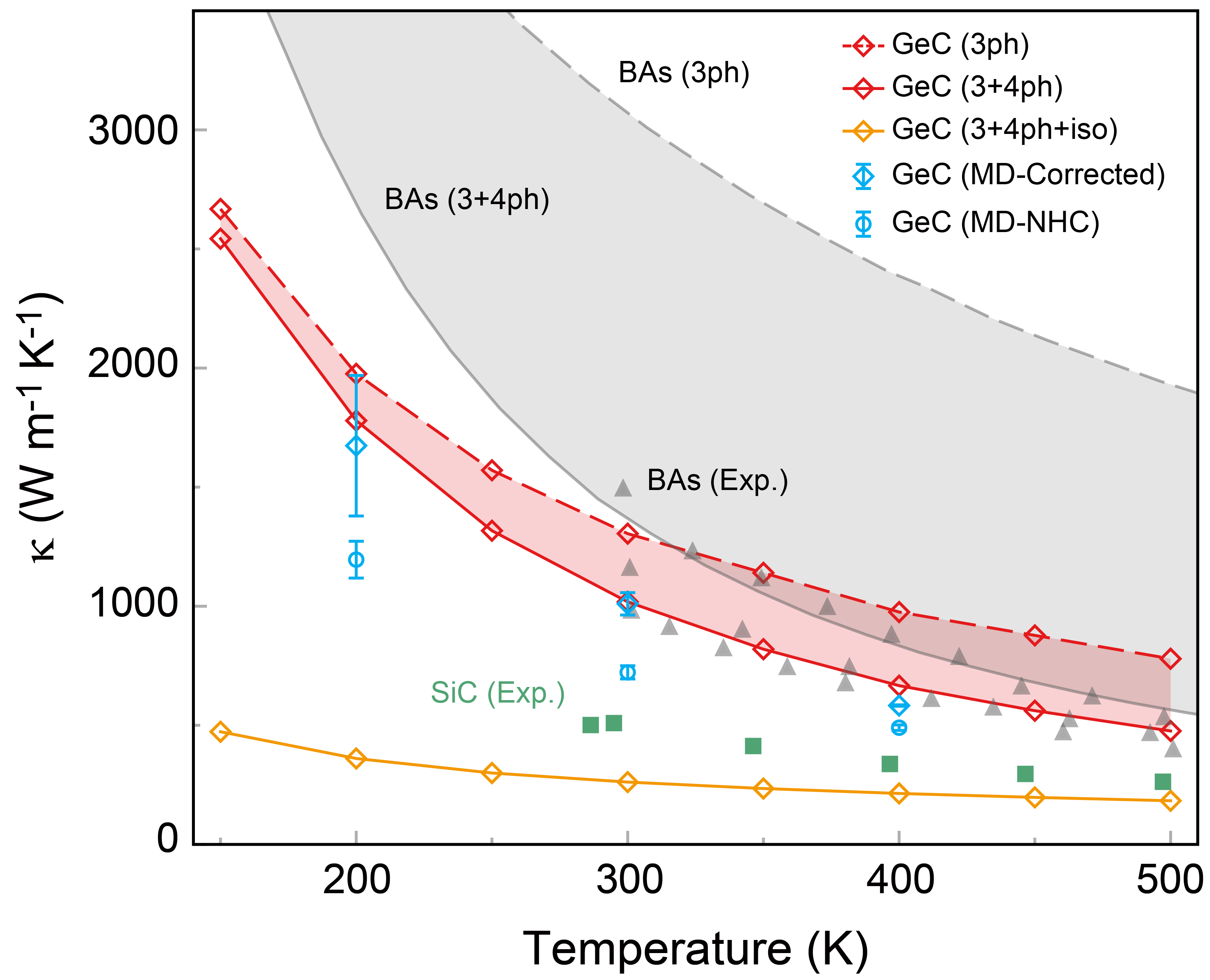}
\caption{\textbf{Thermal conductivity of GeC versus temperature.} Both BTE and MD results are presented, considering different combinations of phonon scattering mechanisms. The MD-NHC results contains the force errors of MLPs, whereas the MD-corrected ones eliminate them via a previous reported extrapolation method~\cite{zhou2025insight,Wu2024}. Also shown are previous 3ph and 3+4ph BTE results for cubic BAs (without isotope scattering)~\cite{Chen2020} and experimental data for BAs~\cite{Tian2018,Li2018,Kang2018} and SiC~\cite{Cheng2022}. The shaded zones highlight the $\kappa$ reductions due to 4ph scatterings.}
\label{fig:fig1}
\end{figure}

To begin with, we examine the overall phonon dispersion of cubic GeC as plotted in Fig.~S1~\cite{SM}. GeC adopts a similar zinc-blende structure like BAs, with one Ge and one C atom per primitive cell. The phonon dispersion has three acoustic branches below about $10$~THz and three optical branches above about $20$~THz, therefore exhibiting a wide a-o gap $\omega_{\mathrm{ao}}$---the hallmark of BAs and $\theta$-TaN. With Slack's conventional criteria~\cite{Slack1973}, the larger mean atomic mass should place $\kappa$ of GeC below even that of cubic silicon carbide (SiC) (around $500~\mathrm{W\,m^{-1}\,K^{-1}}$)~\cite{Cheng2022}. Moreover, with the wide a-o gap in place, GeC is expected to suffer a notable 4ph suppression of $\kappa$ similar to BAs and $\theta$-TaN.

To our surprise, both expectations fail. First, by solving the BTE with 3ph and 4ph scatterings fully included, we observe that isotopically enriched GeC reaches $\kappa=1018~\mathrm{W\,m^{-1}\,K^{-1}}$ at 300~K (Fig.~\ref{fig:fig1}), establishing it as yet another member of the ultrahigh-$\kappa$ family. The substantially higher $\kappa$ compared to SiC can be primarily attributed to the suppression of three-phonon scatterings by the wide a-o gap and bunching of acoustic branches ~\cite{Lindsay2013,Lindsay2013_isotope}. Moreover, we find that 4ph scatterings lower $\kappa$ by merely $22\%$ (down from the 3ph-only value of $1305~\mathrm{W\,m^{-1}\,K^{-1}}$), far less than the $56\%$ penalty in isotopically pure BAs~\cite{Chen2020}. The four-phonon effect weakens further to $10\%$ upon cooling to 200~K. Even at higher temperatures up to 500~K, the 4ph suppression remains at $39\%$.
Across the temperature range of 150 to 500 K, the 3ph-only curve follows the classical relation of $\kappa\propto T^{-1.0}$, whereas the 3+4ph curve steepens only modestly to $T^{-1.4}$, in stark contrast to isotopically enriched BAs with $\kappa\propto T^{-2}$~\cite{Hou2025}. This comparison again indicates surprisingly weak four-phonon interactions. 

We also compute the $\kappa$ of GeC with natural isotope abundance. As shown in Fig.~\ref{fig:fig1}, while the strong mass disorder from the isotope mixtures of Ge and C limits the room-temperature $\kappa$ to $262~\mathrm{W\,m^{-1}\,K^{-1}}$, this value still surpasses about $150~\mathrm{W\,m^{-1}\,K^{-1}}$ for silicon \cite{Glassbrenner1964} and around $170~\mathrm{W\,m^{-1}\,K^{-1}}$ for gallium nitride (GaN) \cite{Yan2024}. Notably, isotope enrichment enhances $\kappa$ of GeC by $289\%$---far exceeding the approximately $50\%$ and $90\%$ enhancement observed in diamond \cite{Onn1992,Wei1993} and cBN \cite{Chen2020}, respectively. To our knowledge, this is the largest isotope effect on $\kappa$ reported to date among materials where 4ph interactions are explicitly accounted for \cite{Lindsay2013_isotope, Yang2019}.

To further validate this anomalously high $\kappa$ and account for phonon scattering non-perturbatively to all orders, we perform machine-learning MD simulations. As isotope scattering is not included in MD, the calculated results correspond to isotopically pure GeC~\cite{Zhou2024mobility,Zhou2024LiH}. While MD simulations of thermal transport have been widely practiced with great success~\cite{Zhou2025superionic,Zhou2026LiNbO3,Zhou2025GaO,yang2026moving,zeng2025lattice}, the raw HNEMD-calculated $\kappa$ values obtained with a global NHC thermostat fall below the first-principles BTE predictions by about $30\%$ (Fig.~\ref{fig:fig1}). 
This underestimation stems from the finite force error of the MLP, which scatters phonons in a manner analogous to isotope disorder~\cite{zhou2025insight,Wu2024}. Eliminating this artifact by extrapolating to the zero-force-error limit based on our aforementioned formula reproduces the BTE results. As exemplified by the 300~K case, this extrapolation yields $\kappa_0 = 1010 \pm 47~\mathrm{W\,m^{-1}\,K^{-1}}$, in good agreement with the BTE prediction of $1018~\mathrm{W\,m^{-1}\,K^{-1}}$. The raw data and the extrapolation procedure are presented in Figs. S8 and S9 of the Supplemental Material~\cite{SM}. The inclusion of higher-order interactions and the excellent agreement between two distinct approaches firmly establish isotope-enriched GeC as a rare material of ultrahigh $\kappa$ above $1000~\mathrm{W\,m^{-1}\,K^{-1}}$ at room temperature.

\begin{figure*}[t]
\centering
\includegraphics[width=0.95\textwidth]{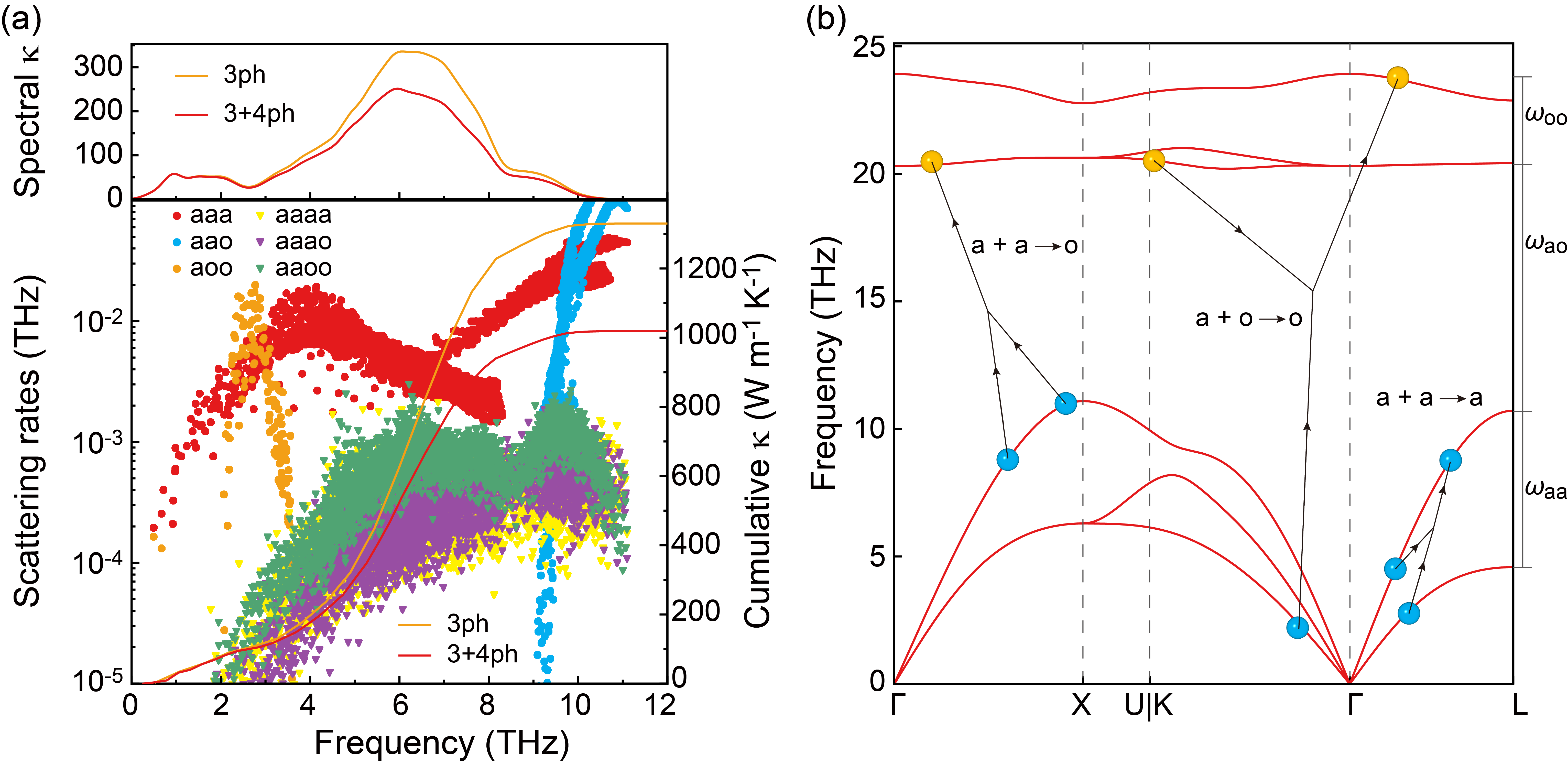}
\caption{\textbf{Spectral analysis and scattering channels in isotope-enriched GeC at 300~K.} (a) Upper panel: spectral $\kappa$ with and without 4ph scatterings. Lower panel: frequency-resolved 3ph and 4ph scattering rates decomposed into the aaa, aao, aoo channels and aaaa, aaao, aaoo channels, respectively, with the overlaid solid curves showing the cumulative $\kappa$ for the 3ph and 3+4ph cases. (b) Phonon dispersion with representative 3ph scattering channels indicated by arrows (a$+$a$\rightarrow$a, a$+$a$\rightarrow$o, and a$+$o$\rightarrow$o). The gaps between the longitudinal and transverse acoustic branches ($\omega_{\mathrm{aa}}$), the acoustic and optical branches ($\omega_{\mathrm{ao}}$), and the longitudinal and transverse optical branches ($\omega_{\mathrm{oo}}$) are marked on the right.}
\label{fig:fig2}
\end{figure*}

To understand the microscopic origin of the weak four-phonon effect in GeC, we first perform spectrally-resolved thermal transport analysis at 300~K (Fig.~\ref{fig:fig2}). The spectral and cumulative $\kappa$ reveal that acoustic phonons below 10~THz are the dominant heat carriers, with the 4-8~THz range alone contributing 75\% of $\kappa$ [Fig.~\ref{fig:fig2}(a)]. The four-phonon suppression is also confined within this window and is much weaker outside it. As a comparison, 4ph scatterings contribute appreciably over the whole acoustic range in BAs~\cite{Ravichandran2019}, as shown in Fig. S2 \cite{SM}. 

Next, we examine the phonon scattering rates, starting with an empirical estimate of the relative strength of 3ph and 4ph processes. In three-dimensional materials, $n$-phonon scattering rate typically scales as $\tau_n^{-1} \sim V_n^2 f^{\,n-2}/\omega^{n+1}$, where $V_n \sim \phi_n/m^{n/2}$ measures the anharmonic coupling through the $n$th-order interatomic force constants $\phi_n$, $f \sim k_{\mathrm{B}}T/\hbar\omega$ is the thermal occupation factor, and $\omega \sim \sqrt{\phi_2/m}$ is the characteristic phonon frequency. In addition, $k_{\mathrm{B}}$ is the Boltzmann constant, $m$ is the atomic mass, and $\hbar$ is the reduced Planck constant. The relative importance of the two processes can thus be evaluated as $\tau_4^{-1}/\tau_3^{-1} \sim (\phi_4/\phi_3)^2 T/\phi_2$  ~\cite{Feng2017, Yue2017}. Substituting the room-temperature force constants of BAs and GeC yields $56$ and $52~\mathrm{K\,eV^{-1}}$, respectively (Table S3~\cite{SM}). 
The close values indicate that anharmonicity of the interatomic forces is not the cause for the weak four-phonon effect in GeC. Rather, the harmonic lattice properties defining the scattering phase space should be primarily responsible.

Subsequently, to reveal which scattering processes are predominant, we decompose the first-principles 3ph scattering rates into the aaa, aao, and aoo channels and the 4ph scattering rates into the aaaa, aaao, and aaoo channels [Fig.~\ref{fig:fig2}(a)]. The ooo and aooo interactions in GeC cannot occur due to the constraint of energy conservation. In addition, the oooo channel does not contribute to $\kappa$, as it involves only optical phonons with near-zero group velocities. Thus, these channels are not shown. Across the entire frequency range, aaa interactions remain strong and mostly dominate over the other channels, despite a dip between 4 and 8~THz that arises from the activation of a selection rule rooted in the bunching of acoustic branches. The aao channel switches on only beyond about 9~THz, the threshold set by $\omega_{\mathrm{ao}}$, while the aoo channel is confined to the 2-4~THz window by $\omega_{\mathrm{oo}}$. In comparison, the 4ph scattering rates present a markedly different picture: the aaaa, aaao, and aaoo processes all persist across the spectrum, with the aaoo channel playing the dominant role. However, even within the dip, the aaa rates remain much higher than those of 4ph scatterings. Moreover, the aao and aoo rates rise sharply within their respective frequency windows, which surpass all the 4ph rates and also aaa scatterings. Their overall predominance contrasts notably with the channel decomposition of BAs (Fig.~S2~\cite{SM}).

To identify the source of this notable difference between GeC and BAs, we analyze their phonon dispersions (Fig.~S3~\cite{SM}), with representative 3ph scattering processes for GeC sketched in Fig.~\ref{fig:fig2}(b). Compared to BAs, GeC has a larger gap between the longitudinal and transverse acoustic (LA and TA) branches ($\omega_{\mathrm{aa}}$), which weakens the bunching effect and relaxes the suppression of aaa scatterings. In addition, the a-o gap ($\omega_{\mathrm{ao}}$) of GeC is about 2\% larger, which tends to suppress the aaoo scatterings, as discussed later. Together, these two factors raise the relative importance of the aaa channel across the spectrum. Of particular interest is the presence of two nearly flat optical branches in GeC---features absent in BAs---which substantially ease the requirements for crystal-momentum conservation and lead to a resonant enhancement of the aao and aoo channels. These results indicate that the synergy between acoustic band bunching and flat optical bands promotes strong three-phonon scatterings, which collectively accounts for the unexpectedly weak four-phonon effect in isotope-enriched GeC.

\begin{figure*}[t]
\centering
\includegraphics[width=\textwidth]{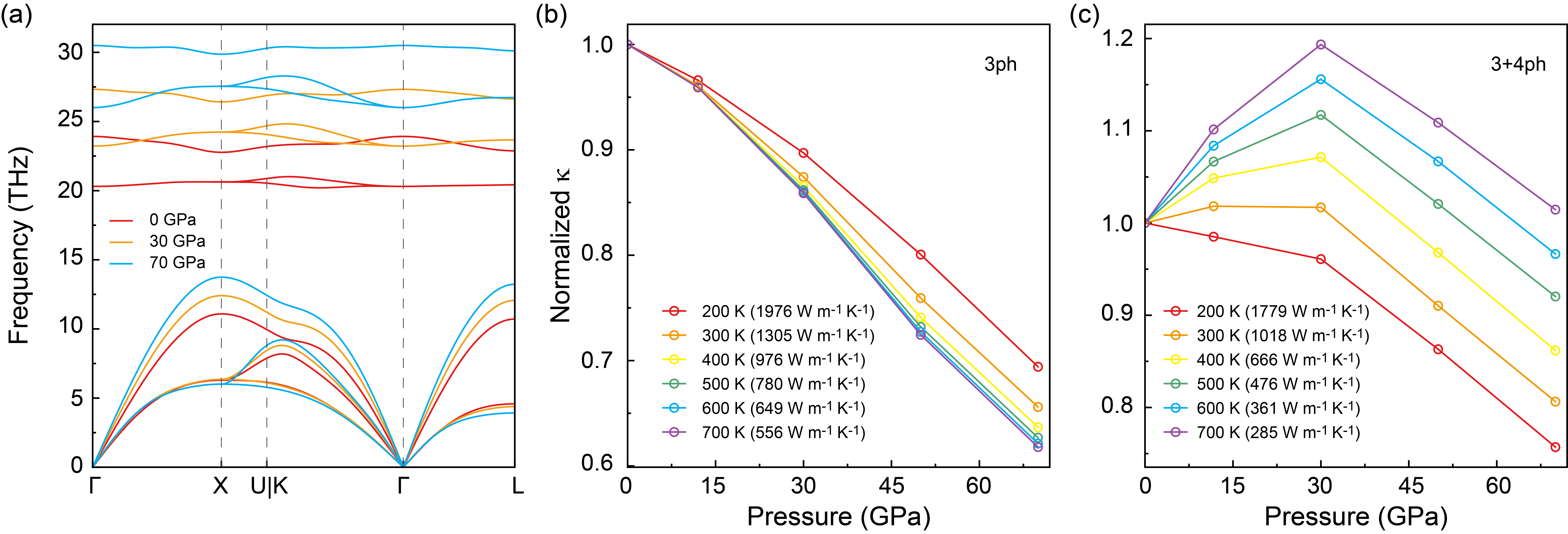}
\caption{\textbf{Pressure-induced anomalous thermal transport in isotope-enriched GeC.} (a) Phonon dispersions under hydrostatic pressures of 0, 30, and 70~GPa at 300 K. (b) Thermal conductivity normalized to its zero-pressure value as a function of pressure at representative temperatures. The corresponding reference values are also given. (c) Same as (b) but with both 3ph and 4ph scatterings.}
\label{fig:fig3}
\end{figure*}

\begin{figure*}[t]
\centering
\includegraphics[width=\textwidth]{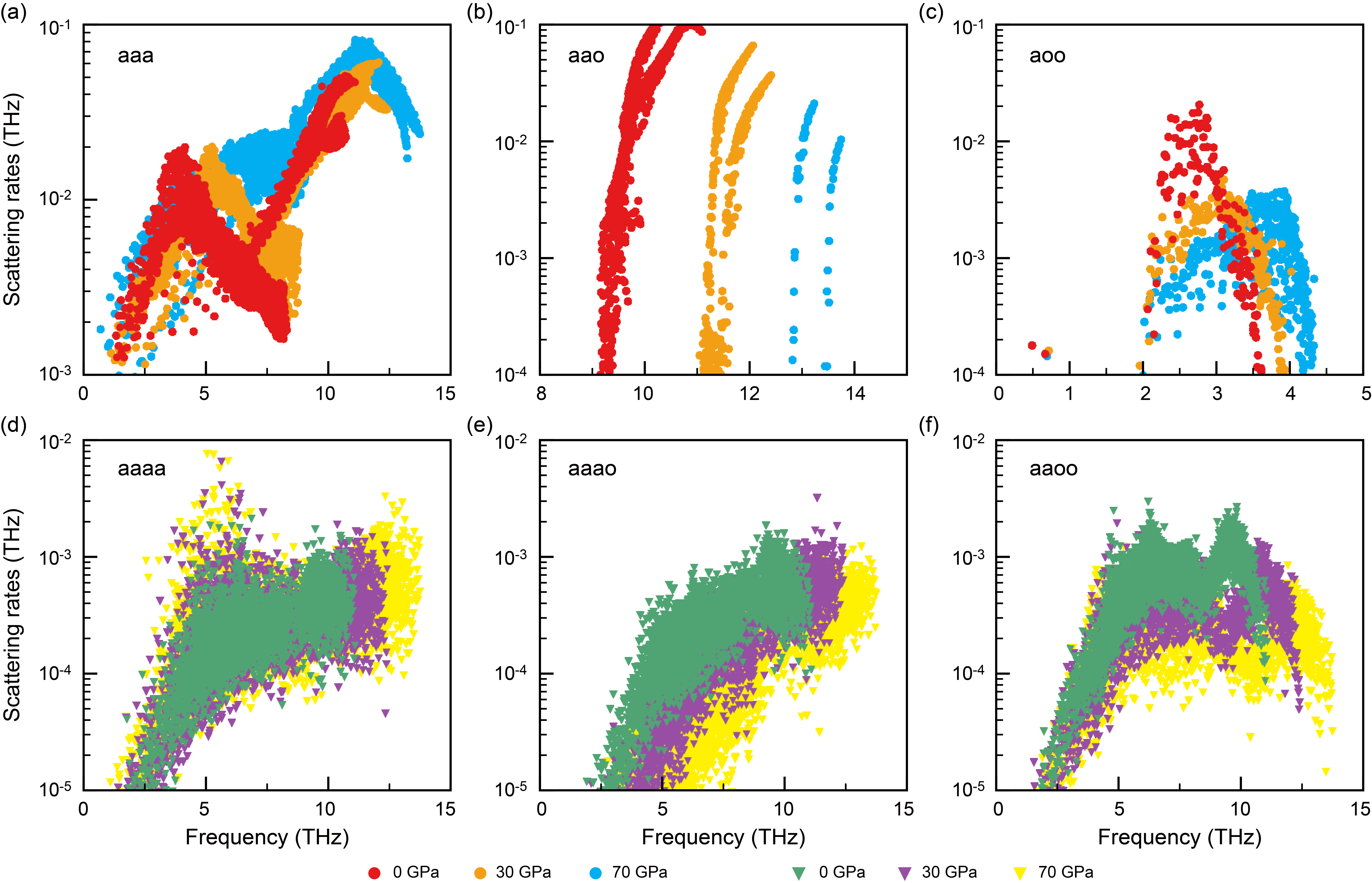}
\caption{\textbf{Channel-resolved phonon scattering rates in isotope-enriched GeC under different pressures at 300~K.} The 3ph interactions are decomposed into the aaa, aao, aoo channels, while the 4ph processes include the aaaa, aaao, aaoo channels.}
\label{fig:fig4}
\end{figure*}

The anomalous thermal transport properties of GeC under under ambient conditions motivates us to further investigate the impact of an external hydrostatic pressure, which serves a powerful degree of freedom for tuning the phonon bands~\cite{Ravichandran2019,Li2022,Kundu2024,dai2022simultaneous}. Traditionally, the effect of pressure on $\kappa$ is described by the Leibfried-Schl\"omann formula, $\left(\partial\ln\kappa/\partial P\right)_{T} = g/B_{\mathrm{T}}$, where $g$ is a constant and $B_{\mathrm{T}}$ is the isothermal bulk modulus. This relation predicts a monotonic dependence. Recently, pressure-induced non-monotonic thermal transport in BAs~\cite{Ravichandran2019,Li2022} and $\theta$-TaN~\cite{Kundu2024} have been reported. Below, we explore the possibility of non-trivial pressure dependence in isotope-enriched GeC.

In Fig.~\ref{fig:fig3}(a), we plot the phonon dispersions of GeC under three representative pressures including 0, 30, and 70 GPa. All three optical branches and also the LA branch stiffen with rising pressure. The two TA branches, by contrast, generally soften near certain high-symmetry points such as X and L. Since the optical branches are more sensitive, the a-o gap $\omega_{\mathrm{ao}}$ widens. In addition, the flat TO branch becomes more dispersive at higher pressures, which reduces the resonance between acoustic and optical phonons. Meanwhile, the group velocities of the LA phonons increase with \mbox{pressure}. Therefore, one would naturally expect the $\kappa$ of GeC to increase with pressure if variations in phonon scattering rates can be ignored. However, 3ph-only BTE calculations predict the opposite, with $\kappa$ decreasing monotonically with pressure at all temperatures from 200 to 700 K [Fig.~\ref{fig:fig3}(b)]. In Fig.~\ref{fig:fig4}, we analyze and compare the channel-resolved 3ph scattering rates under different pressures. Two competing effects are at play here. First, as pressure increases, the larger $\omega_{\mathrm{ao}}$ shifts the aao processes toward higher frequencies. Moreover, both the aao and aoo channels are suppressed due to the weakening of acoustic-optical resonance. These variations in phonon scattering tend to increase $\kappa$. However, the widening LA-TA gap $\omega_{\mathrm{aa}}$ with rising pressure diminishes the bunching effect and activates the aaa channel, which ultimately dictates the total phonon scattering rates (Fig.~S4~\cite{SM}) and therefore leads to decrease of $\kappa$.

Intriguingly, the dependence of $\kappa$ on pressure becomes more complex once 4ph scatterings are included. As shown in Fig.~\ref{fig:fig3}(c), $\kappa$ still decreases monotonically with pressure at 200~K but at reduced rate. At higher temperatures, however, $\kappa$ varies non-monotonically---first increasing and then decreasing. For instance, at 700~K, $\kappa$ peaks at around 30 GPa, surpassing the ambient value by $19\%$. This anomalous pressure-induced thermal transport in GeC can also be understood based on the channel-resolved and total scattering rates in Fig.~\ref{fig:fig4} and Fig.~S4~\cite{SM}, respectively. Unlike the 3ph aaa channel, 4ph scatterings generally weaken with increasing pressure, especially for the dominant aaoo channel, since the larger $\omega_{\mathrm{ao}}$ progressively restricts the available phase space. At sufficiently low temperatures, the 4ph contribution becomes too weak to counterbalance the 3ph-induced decline in $\kappa$. In contrast, at elevated temperatures with comparable 4ph and 3ph interactions, the pressure-induced suppression of 4ph scatterings initially outweighs the increase in 3ph scatterings, which explains the rise of $\kappa$. As pressure further increases, the 4ph channels become negligible and the 3ph processes again take over, therefore leading to the decrease in $\kappa$.

In summary, we theoretically predict a thermal conductivity of $1018~\mathrm{W\,m^{-1}\,K^{-1}}$ at room temperature, even with higher-order phonon scatterings fully incorporated. This remarkable property originates from two distinct mechanisms: an extraordinary isotopic enhancement effect of 289\% and a surprisingly modest suppression of about 22\% by four-phonon scattering---despite the presence of a wide acoustic–optical phonon gap. We reveal that this protection from 4ph scattering is rooted in the synergy between the flat optical bands and bunched acoustic branches, which collectively sustain 3ph scatterings at a relative strong level. We further demonstrate that the $\kappa$ of GeC varies non-monotonically with external hydrostatic pressure, deviating from the monotonic trend predicted by the classical Leibfried-Schl\"omann formula. Notably, these features of GeC also suggest a potential for high ambipolar mobility, reminiscent of that observed in BAs \cite{liu2018simultaneously}, which lies beyond the scope of the present work. Altogether, our findings establish isotope-enriched cubic GeC as an intriguing platform for studying exotic phonon physics and a member of the ultrahigh-thermal-conductivity family, offering a route toward the effective dissipation of extreme heat fluxes in advanced electronics.

\FloatBarrier

\begin{acknowledgments}
This work was financially supported by the Science Fund for Creative Research
Groups from the National Natural Science Foundation of China (No.~52521007),
the Scientific Research Innovation Capability Support Project for Young
Faculty (ZYGXQNJSKYCXNLZCXM-E1) from the Ministry of Education of China, the
National Key R\&D Program of China (No.~2024YFA1207900). W.Z.~acknowledges
support from the National Natural Science Foundation of China (No.~525B2087)
and China Association for Science and Technology. B.S.~acknowledges support
from the New Cornerstone Science Foundation through the XPLORER PRIZE. We
appreciate the High-performance Computing Platform of Peking University for
supporting our simulations.
\end{acknowledgments}

\bibliography{GeC}

\end{document}